\documentclass[notitlepage,nofootinbib]{revtex4-1}
\usepackage{ulem}
\usepackage{amssymb, amsmath, amsopn, color, graphicx}
\usepackage[english]{babel}
 
\usepackage{bm}
\DeclareMathAlphabet{\matheul}{U}{eus}{m}{n}
\usepackage{float}

\begin{document}


\title{A curious general relativistic sphere}
\thanks{}%

\author{Philip Beltracchi and Paolo Gondolo}
\email{phipbel@aol.com, paolo.gondolo@utah.edu}
\affiliation{Department of Physics and Astronomy, University of Utah, 115 South 1400 East Suite 201, Salt Lake City, UT 84012-0830}
 
\collaboration{}

\date{\today}
 
\begin{abstract}
We present a particular solution to the Einstein field equations that could arise as a magnetic monopole in the strong field limit of the Born-Infeld electrodynamics.  We solve its geodesics exactly and show that they can mimic the gravitational lensing of a singular isothermal sphere, but with matter and radiation following the same trajectories. This solution has previously been called ``cloud of strings'' or ``string hedgehog,'' and has appeared as the limit of a global O(3) monopole.
\begin{description}
\item[Usage]
 
\item[PACS numbers]
 
\item[Structure]
\end{description}
\end{abstract}
 
\pacs{Valid PACS appear here}
\maketitle
 

\section{Introduction}
We examine the properties of a particular solution to the Einstein field equations that could arise as a magnetic monopole in the strong field limit of the Born-Infeld electrodynamics.  It can mimic the gravitational lensing of a singular isothermal sphere. Its geodesics are exactly solvable, showing that matter and radiation follow the same trajectories.

The history of this solution is somewhat complicated and features rediscoveries arising from the different physical realizations of the solution. 

Letelier \cite{Letelier:1979ej} discovered the koosh solution when looking for the gravitational fields created by fluids made of classical strings, but did not study their geodesics. 
Barriola and Vilenkin \cite{PhysRevLett.63.341} wrote the koosh metric as a large $r$ limit of their O(3) monopole, and discussed gravitational lensing in analogy to cosmic strings. 

Guendelman and Rabinowitz \cite{Guendelman:1991qb} showed that the koosh metric appears in the $\sigma$-model limit of the O(3) monopole (more precisely: a field $\phi^a$, $a=1,2,3$, with only kinetic terms in the Lagrangian, but confined to a sphere $\phi^a\phi^a=v^2$ in field configuration space). They studied the monopole configuration $\phi^a = \pm v (r^a/r)$, found that it has what we call  uniaxial stress-energy with $p_T =0$ and $p_r=-v^2/r^2$). Guendelman and Rabinowitz also rediscovered the string configuration of Letelier without seemingly being aware of Letelier's paper. They also wrote about the superposition of uniaxial solutions, without reference to~\cite{Gurses:1975vu}, and examined the effects of superposing koosh-like solutions to vacuum bubbles. 

Spacetimes consisting of two separate string cloud regions  have been analyzed in \cite{Delice:2003zp,HabibMazharimousavi:2017zlc}, 
with attention to the properties of the boundary between the two regions. 

Geodesics of the Barriola-Vilenkin global monopole and of the Schwarzschild black hole with string cloud backgrounds have been studied in \cite{Shi:2009nz} and \cite{2017IJMPD..2641005B}, respectively. 

We show that both null and nonnull geodesics of the plain koosh have interesting properties, and compare them to the geodesics of an isothermal sphere and to astrophysical observations of dark matter halos. We also add a connection to nonlinear electrodynamics, in particular the Born-Infeld magnetic monopole.

\section{Einstein equations}
 In this paper we consider spherically symmetric systems with energy-momentum tensors of Segre type [(11)(1,1)] (see \cite{Stephani:2003tm}). Because these systems have vacuum properties of 1) $p_i=-\rho$, 2) boosts symmetry along, and 3) rotation symmetry about one axis at each point in spacetime, we call them uniaxial in analogy with the terminology from liquid crystals. We examine some general considerations of spherically symmetric uniaxial systems briefly here.
 
 Using spherical coordinates $(t,r,\theta,\phi)$, where $r$ is the area radius, a general static spherically symmetric metric has a line element
\begin{equation}
   ds^2= - e^{2\Phi(r)} \,  dt^2 + \frac{dr^2}{1-\frac{2m(r)}{r}} + r^2 \, d\theta^2 + r^2 \, \sin^2(\theta) \, d\phi^2. 
   \label{met1}
\end{equation}
The corresponding energy momentum tensor is
\begin{align}
    T^\nu_{~\mu} = \, \begin{pmatrix}
-\rho & 0 & 0 & 0 \\
0 & p_r & 0 & 0 \\
0 & 0 & p_T & 0 \\
0 & 0 & 0 & p_T 
\end{pmatrix}
\end{align}
Here $\rho$ is the energy density, $p_r$ is the pressure in the radial direction, and $p_T$ is the pressure in the tangential directions.
 Spherical symmetry and the Segre type [(11)(1,1)] requires $p_r=-\rho$. Einstein's equations then give
\begin{align}
     \frac{dm}{dr}= 4\pi r^2 \rho,
   \label{mdefined}\\
   \frac{d\Phi}{dr}= \frac{1}{2}\frac{d}{dr}\log\left(1-\frac{2m}{r}\right),
   \label{eeqpotential}\\
 \frac{r}{2} \frac{d \rho}{dr}+\rho=-p_T.
   \label{speacialTOV}
\end{align}
 Equation~(\ref{eeqpotential}) can be integrated to give $\Phi$. Re-scaling the $t$ coordinate to absorb the constant of integration, we can write the metric with $-g_{tt}$ and $g_{rr}$ as inverses and obtain
 \begin{equation}
   ds^2=-\left(1-\frac{2  m(r)}{r}\right) dt^2+\left(1-\frac{2 m(r)}{r}\right)^{-1}dr^2+ r^2 d\theta^2+r^2 \sin^2\theta \, d\phi^2.
   \label{generalkerrschildmet}
\end{equation}
The metric in  Eq.~(\ref{generalkerrschildmet}) is of the Kerr-Schild class \cite{Dymnikova2018}, which means it can be written $g_{\mu \nu}=\eta_{\mu \nu}-S k_\mu k_\nu$, where $S$ is a scalar function and $k^\mu$ is a null vector with respect to both $g_{\mu \nu}$ and $\eta_{\mu \nu}$ \cite{Stephani:2003tm}. We can introduce the new coordinate $t^*$, satisfying
 \begin{equation}
     dt^*=dt+dr-\left(1-\frac{2 m(r)}{r}\right)^{-1}dr,
 \end{equation}
 and obtain
 \begin{equation}
     ds^2=-d t^{*2}+dr^2+r^2 d\theta^2+r^2 \sin^2\theta d\phi^2+\frac{2 m(r)}{r}(dt^*-dr)^2.
     \label{KSformmet}
 \end{equation}
 The first four terms are the Minkowski metric in spherical coordinates, $S(r)=2m(r)/r$ and $k_\mu dx^\mu=dt^*-dr$.
 It is known that Kerr-Schild systems sharing a null vector\footnote{Interestingly a similar behavior exists even if one of the systems is not uniaxial/Kerr-Schild. Suppose we have one spherical uniaxial system with $T^{(1)}_{\mu\nu}$ and a second concentric system with $T^{(2)}_{\mu\nu}$ and both energy-momentum tensors are separately conserved. This conservation results in
\begin{align}
    \begin{split}
    \frac{dp^{(1)}_r}{dr}= - \frac{ \left( m+4\pi r^3( p^{(1)}_r +p^{(2)}_r)\right) }{r^2 \left( 1 - \frac{2  m}{r} \right) }\left( \rho^{(1)}+p^{(1)}_r \right)+ \frac{2(p^{(1)}_T-p^{(1)}_r)}{r},\\
    \frac{dp^{(2)}_r}{dr}= - \frac{ \left( m+4\pi r^3( p^{(1)}_r +p^{(2)}_r)\right) }{r^2 \left( 1 - \frac{2  m}{r} \right) }\left( \rho^{(2)}+p^{(2)}_r \right)+ \frac{2(p^{(2)}_T-p^{(2)}_r)}{r}
    \label{TOVA2}
    \end{split}
\end{align}
Note importantly that since system 1 is of Segre type $[(11)(1,1)]$, the factor $\rho^{(1)}+p^{(1)}_r =0$ and the gravitational force term acting on the uniaxial system 1 drops out. This means the density profile of system 1 can be unchanged by the presence of system 2 if spherical symmetry is maintained. System 2 is generally affected by system 1 due to the contributions of system 1 to the gravitational potential.} may be superimposed \cite{Gurses:1975vu}. In practice, for spherically symmetric uniaxial systems, one superimposes by simply using a metric of the form Eq.~(\ref{generalkerrschildmet}) and adding the $m$ functions from the constituent systems. One does not have to stop at two systems. Superimposing multiple $m$ functions has been considered \cite{Gibbons:1992gt, 2017IJMPD..2641005B,Secuk:2019njc}. 

In addition to simple superposition properties, uniaxial systems may be joined with simple junction layers.
Suppose we have a system consisting of an interior region and an exterior region, both of which are uniaxial and described by a Kerr-Schild metric written in the form of Eq.~(\ref{generalkerrschildmet}), joined at a hypersurface with unit normal $n^\alpha$ (which we assume not null). Let $e^\alpha_a$  be a basis of vectors tangent to the hypersurface, and let $[a]$ denote the jump of $a$, defined as $[a]=a_{\rm(out)}-a_{\rm(in)}$, the difference of the values of $a$ on ``inner'' and ``outer'' side of the hypersurface.  The junction conditions (see e.g.~\cite{poissonAGR}) require that the induced metric on the junction hypersurface $h_{ab}=g_{\alpha \beta}e^\alpha_a e^\beta_b$ be the same on both sides of the surface, i.e., $[h_{ab}]=0$, and that the discontinuity in extrinsic curvature of the hypersurface $K_{ab}=(\nabla_\beta n_\alpha) e^\alpha_a e^\beta_b$ be related to the surface stress-energy by 
\begin{align}
    S_{ab}=\frac{1}{8\pi}([K_{ab}]-[K]h_{ab}) , \\
    T^{\alpha \beta}_{\rm(surface)}=\delta(l)S^{ab}e^\alpha_a e^\beta_b.
\end{align}
Here $l$ is proper distance from the hypersurface along its normal. For a hypersurface defined by $r=R$, we can choose coordinates $y^a$ on the hypersurface such that the $e^\alpha_a$ components become 0 or 1 and
\begin{equation}
    h_{ab}dy^a dy^b=g_{tt}dt^2+R^2 d\theta^2+R^2 \sin(\theta)^2 d\phi^2.
\end{equation}
The first junction condition $[h_{ab}]=0$ is then satisfied as long as $g_{tt\rm(in)}=g_{tt\rm(out)}$, which requires the function $m$ be continuous at $R$. The second junction condition requires that the surface stress-energy tensor is
\begin{equation}
T^2_{~2\rm(surface)}=T^3_{~3\rm(surface)}=\frac{R}{2}\, [\rho(R)] \, \delta(r-R)
\label{juntion2}
\end{equation}
with the other surface stress-energy terms being 0. Here $[\rho(R)]$ is the jump of energy density at the hypersurface. Note that the form of the surface stress in Eq.~(\ref{juntion2}) is a $p_T$ type term. Also note that if the density is continuous, there is no stress-energy associated with the hypersurface. This means that in addition to superimposing uniaxial systems one may easily define systems with separate regions, at least in spherical symmetry.
 \section{Solution for $p_T=0$}
One way to differentiate between spherically symmetric uniaxial systems is to specify an equation of state for $p_T$.  For this paper we use
\begin{equation}
   p_T=0.
   \label{ourEOS}
\end{equation}
 With this equation of state, plus the condition $\rho=-p_r\ge0$, the null, weak, strong, and dominant energy conditions are all satisfied since $|p_r|\le \rho$, $|p_T|\le \rho$, $\rho+p_r+2p_T\ge0$. Zero is the lowest value $p_T$ can have and still satisfy the strong energy condition. The density profile dictated by Eqs.~(\ref{speacialTOV}) and (\ref{ourEOS}) is
 \begin{equation}
    \rho=\frac{\lambda }{4\pi r^2},
    \label{uniaxial density}
 \end{equation}
 where $\lambda$ is a constant with units mass per length. If we assume $M=0$ (there is no ``point mass" at the origin), Eq.~(\ref{mdefined}) gives
 \begin{equation}
    m=\lambda r,
    \label{uniaxialmass}
 \end{equation}
 and the metric becomes
 \begin{equation}
     ds^2=-\kappa^2  dt^2+\frac{1}{\kappa^2}dr^2+ r^2 d\theta^2+r^2 \sin^2\theta \, d\phi^2,
   \label{uniaxialmet}
 \end{equation}
where $\kappa=\sqrt{1-2  \lambda}$. Notice that the $g_{tt}$ and $g_{rr}$ metric functions are constant everywhere. This metric describes a hypercone in four dimensions in that the circumference of a circle of proper radius $r^*$ is $2\pi \kappa r^*$. Its Riemann tensor has only one independent nonzero element $R^{\theta \phi}_{~~\theta \phi}=(1-\kappa^2)/r^2$. Demanding that the $t$ coordinate remains timelike requires 
 $ 2\lambda<1.$ For notation we refer to this as a ``koosh/monopole" solution because it can arise from monopole solutions or from a very large number of strings going through a point like a koosh ball. This solution was found in~\cite{Letelier:1979ej}, where it was called a ``cloud of strings'', in~\cite{PhysRevLett.63.341}, as an asymptotic limit of a global O(3) monopole, and in~\cite{Guendelman:1991qb}, as a cloud of strings (which they call a ``string hedgehog'') and as a particular O(3) monopole (which they call a ``hedgehog'').
 
\section{Geodesics}
The geodesics obey the equation
\begin{equation}
   g_{\mu \nu}\frac{\partial x^\mu}{\partial \tau}\frac{\partial x^\nu}{\partial \tau}=-\epsilon,
   \label{geo1}
\end{equation}
where $\epsilon=1$ and $\tau$ is proper time for timelike geodesics, and $\epsilon=0$ and $\tau$ is an affine parameter for null geodesics. Because of the spherical symmetry and static metric, any plane through the origin is equivalent so we describe trajectories in the $\theta=\pi/2$ equatorial plane. A further consequence of the symmetry is the existence of conserved quantities
$L= r^2(d\phi/ d\tau)$ and $E=\kappa^2(dt /d\tau).$
Equation~(\ref{geo1}) becomes 
\begin{equation}
 E^2=\left(\frac{dr}{d\tau}\right)^2-V_{\rm eff}(r), 
 \label{Veff}
\end{equation}
with the effective potential $V_{\rm eff}(r)=\kappa^2\left(\epsilon+L^2/r^2\right)$. There is no minimum of $V_{\rm eff}$ so there are no bound orbits. We can eliminate $\tau$ and obtain a differential equation for the trajectories
\begin{equation}
   \left(\frac{du}{d\phi}\right)^2=\frac{E^2-\kappa^2(L^2 u^2 +\epsilon)}{L^2},
   \label{phivar1}
\end{equation}
where $u=1/r$. It is useful to take a $\phi$ derivative of Eq.~(\ref{phivar1}), which gives
\begin{equation}
   \frac{d^2 u}{d\phi^2}=-\kappa^2 u.
   \label{d2trajks}
\end{equation}
 Equation~(\ref{d2trajks}) is a simple harmonic oscillator equation and can be easily solved to give trajectories
\begin{equation}
 r \cos(\kappa \phi)=\kappa b,
 \label{traj}
\end{equation}
where $b$ is the impact parameter.

\begin{figure}
   \centering
   \includegraphics[width=8cm]{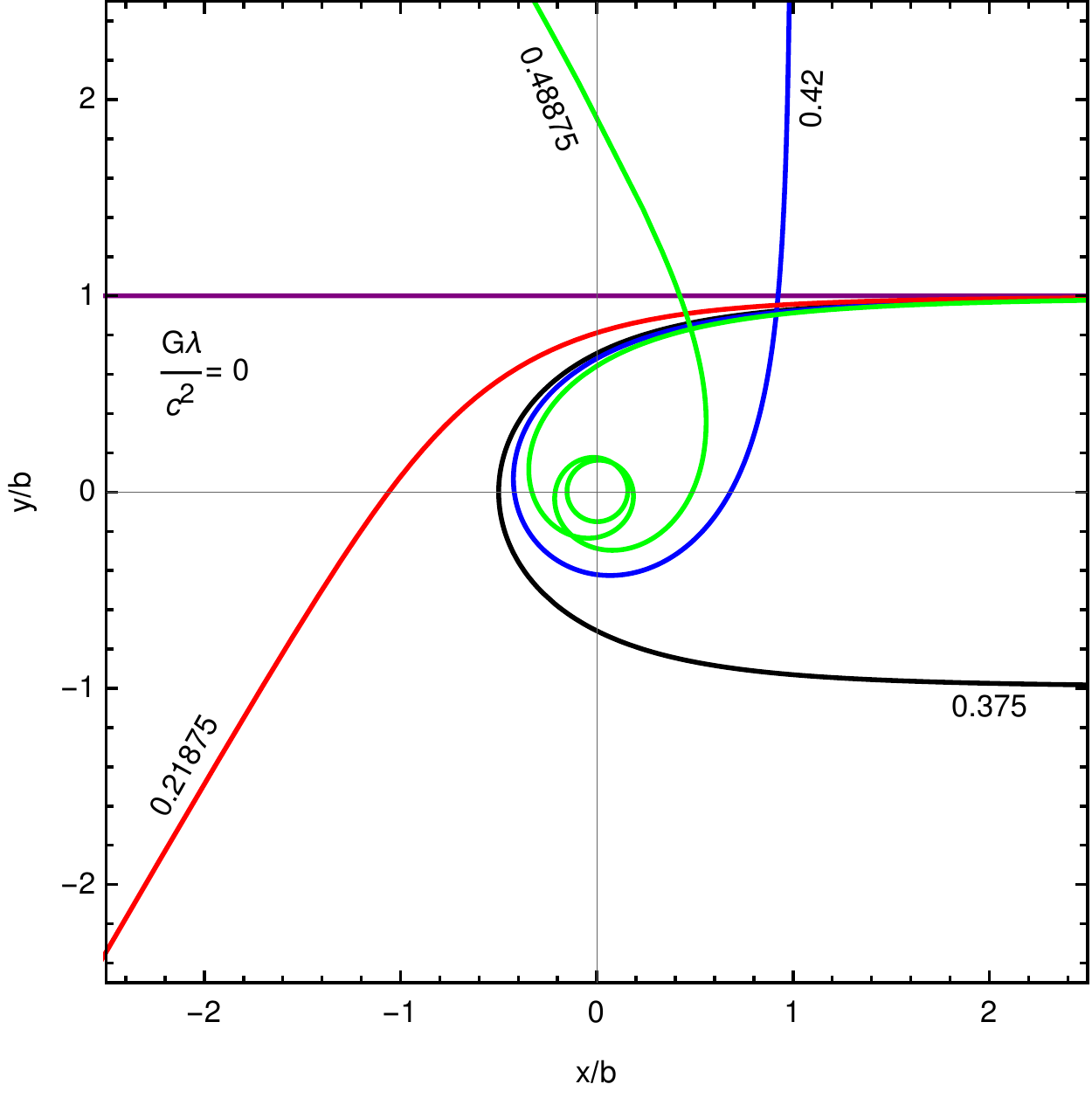}
   \caption{Trajectories of geodesics in the equatorial plane for various values of $\lambda$. The shape of trajectories is independent of $\epsilon$, and the impact parameter $b$ only acts as a scaling factor. Note that as $\lambda$ increases the trajectories become more bent.}
   \label{trajectories}
\end{figure}

There are two special properties of trajectories for this solution. The first is that $\epsilon$ drops out, so lightlike and massive particles follow the same trajectories. The second is that the deflection angle
\begin{equation}
   \alpha=\pi \frac{1-\kappa}{\kappa}
   \label{angle}
\end{equation}
is independent of the impact parameter. This independence occurs in vacuum cosmic string solutions \cite{PhysRevD.23.852,Gott:1984ef,Kotvytskiy:2015sqa} and in singular isothermal spheres. When $\kappa$ goes to 1 we recover Minkowski space, Eq.~(\ref{traj}) describes a straight line, and Eq.~(\ref{angle}) dictates the deflection angle is 0. For $\alpha>\pi$, trajectories wrap around the origin and intersect themselves. This happens when 
$\kappa<1/2$ or equivalently $\lambda>3/8.$ Figure \ref{trajectories} shows some examples of trajectories.
\section{An asymptotically de Sitter combination}
Because of the superposition behavior of systems with uniaxial symmetry, it is prudent to at least examine a combined system of a koosh/monopole with a point mass $M>0$ and cosmological constant $\Lambda>0$. (Other sorts of uniaxial combinations featuring a koosh/monopole term have been considered, see e.g. \cite{Gibbons:1992gt, 2017IJMPD..2641005B,Secuk:2019njc}). Then
\begin{equation}
    -g_{tt}=\frac{1}{g_{rr}}=\left(1-\frac{2M}{r}-2\lambda-\frac{\Lambda r^2}{3}\right).
    \label{combinedgtt}
\end{equation}
This system has a Schwarzschild horizon at small $r$ and de Sitter horizon at large $r$. The $r\rightarrow0$ and $r\rightarrow\infty$ behavior is dominated by the Schwarzschild and de Sitter terms, respectively. The presence of a koosh/monopole causes the Schwarzschild horizon to move outward (to $R=2M/\kappa^2$ if $\Lambda$ is neglected) and the de Sitter horizon to move inward (to $R=\kappa \sqrt{3/\Lambda}$ if $M$ is neglected).

 Concerning orbits, Eq.~(\ref{Veff}) applies with $E=-g_{tt}(dt/d\tau)$ and
\begin{equation}
  V_{\rm eff}=  \left(\kappa^2-\frac{2M}{r}-\frac{\Lambda r^2}{3}\right)\left(\frac{L^2}{r^2}+\epsilon\right).
  \label{veffcombined}
\end{equation}
This potential has bound orbits, as follows from studying the sign of $dV_{\rm eff}/dr$. Circular orbits occur at radii $r$ corresponding to an equivalent Schwarzschild-de Sitter solution with mass $M/\kappa^2$ and cosmological constant $\Lambda/\kappa^2$.
\section{Discussion}
There are other systems where the mass increases linearly with radius like  Eq.~(\ref{uniaxialmass}),  and it is useful to draw comparisons.

It is intriguing that if one were to truncate a koosh/monopole at radius $R$ with an appropriate surface junction layer and embed it in a vacuum, the total mass of the system would be $M=\lambda R$. In the limit $\kappa\rightarrow 0^+$, equivalently $\lambda\rightarrow0.5^-$, we get $2M=R$, which is the horizon radius of a Schwarzschild black hole \footnote{Since our first version of this paper, the $\lambda\rightarrow0.5^-$ limit has been studied in more detail, see \cite{Lemos:2020hsz}}. Additionally, one could replace the central singularity with a microscopic de Sitter core, or embed a koosh/monopole in de Sitter space  (or a Reissner Nordstrom exterior) without any junctions if the density is continuous.
 
One system well known for having a density proportional to $1/r^2$ is the singular isothermal sphere. In particular, the Newtonian singular isothermal sphere has a mass profile defined in terms of a velocity dispersion $\sigma_v$ as
\begin{equation}
   m(r)=2\sigma_v^2 r.
\end{equation}
This corresponds to Eq.~(\ref{uniaxialmass}) when $2\sigma_v^2=G\lambda$. The deflection angle for singular isothermal spheres in gravitational lensing is 
\begin{equation}
   \alpha=4\pi \sigma_v^2,
\end{equation}
which is independent of the impact parameter. In the low $\lambda$ limit, the deflection angle for  Eq.~(\ref{angle}) becomes
\begin{equation}
   \alpha\approx \pi \lambda.
\end{equation}
Thus the deflection angle of a koosh/monopole is equivalent to that of a Newtonian isothermal sphere that is half as dense. Therefore, if a koosh/monopole was observed in a lensing survey, it would appear the same as a galaxy, without containing baryonic or dark matter. Unlike a standard dark matter halo, a koosh/monopole Eq.~(\ref{uniaxialmet}) does not, by itself, support bound orbits. Adding $M$ and $\Lambda$ as in Eq.~(\ref{combinedgtt}) does not give flat rotation curves either.
 
Another system with a density profile of the form $1/r^2$ is a TOV star with a linear equation of state $p=k \rho$  \cite{1969Afz.....5..223B,1970ApJ...160..875B,1972grec.conf..185C,Chavanis:2007kn}. Because of its equation of state and density profile, this system is sometimes called a general relativistic isothermal sphere. The mass function and $g_{rr}$ metric function are the same as for a koosh/monopole, with $\lambda=2k/(1+6k+k^2)$. The largest possible $\lambda$ for general relativistic singular isothermal spheres is $\lambda=1/4$ when $k=1$, which is half the value a koosh/monopole takes in the $\kappa\rightarrow0^+$ limit. Also, the $g_{tt}$ metric function is a nonzero power of $r$ (spherical metrics which approach $g_{tt}$ a power of $r$ and constant $g_{rr}\ne1$ are referred to as asymptotically conical in \cite{Cvetic:2012tr}) which is different than a koosh/monopole, because Einstein's equations only give Eq.~(\ref{eeqpotential}) when $p_r=-\rho$.
 
 Many uniaxial spherical objects arise in nonlinear electrodynamics theories \cite{Lobo:2006xt, Gibbons:1995cv}, and are commonly used to construct nonsingular black holes \cite{AyonBeato:1998ub,Elizalde:2002yz, Chinaglia:2017uqd,Fan:2016hvf,Balart:2014cga,Rodrigues_2018}. We find the $p_T=0$ system has the same energy-momentum tensor as a magnetic monopole  in a theory with the Lagrangian 
\begin{equation}
   {\cal L}=-b\sqrt{\frac{F^{\mu \nu}F_{\mu\nu}}{2}},
   \label{ourlag}
\end{equation}
 where $F^{\mu \nu}$ is the electromagnetic field strength tensor. Eq.~(\ref{ourlag}) is the strong magnetic field limit to the famous Born-Infeld \cite{Born:1934gh} Lagrangian 
 \begin{equation}
 {\cal L}= b^2 \left(1-\sqrt{1+\frac{F^{\mu \nu}F_{\mu\nu}}{2 b^2}-\left(\frac{\tilde{F}^{\mu \nu}F_{\mu \nu}}{4 b^2}\right)^2}\right)
 \end{equation}
 with $\tilde{F}^{\mu \nu}=\epsilon^{\mu \nu \alpha \beta}F_{\alpha \beta}/2$ (for magnetic monopoles the term with $\tilde{F}^{\mu \nu}$ is zero). Although the Born-Infeld theory was originally proposed to regularize the self-energy of electric point charges, it also arises in string theory as the effective action for gauge fields on a D-brane \cite{Leigh:1989jq}. It was previously noted \cite{Gibbons:1995cv} that under certain circumstances Born-Infeld solutions lead to a conical singularity and $\rho\propto 1/r^2$ as $r\rightarrow0$. If we were to consider the source of $T_{\mu \nu}$ as a magnetic monopole, it would be natural to add a point mass term to the $m$ function, making the object a black hole. 
 
 One final point worth considering is if there were multiple nonconcentric systems resembling a koosh/monopole. One possibility is that the uniaxial character of the total energy-momentum tensor would be maintained, and there would be a well-defined axis at every point along which $p=-\rho$.  One could think of this as strings which line up smoothly or nonlinear magnetic field lines. Another possibility is that the superposition would lead to an isotropization of the pressure due to axes nearby pointing in many directions, as in tangled strings, leading to a net value of $p=-\rho/3$, like in the coasting cosmology \cite{1989ApJ...344..543K}.
 \section{Acknowledgements}
 We would like to thank Jose Pizarro de Sande e Lemos and Gary Gibbons for bringing references \cite{Letelier:1979ej,PhysRevLett.63.341,Cvetic:2012tr} to our attention. This work was partially supported by NSF award PHY-1720282 at the University of Utah.
 
\bibliographystyle{apsrev4-1}
\bibliography{72320.bib}

\end{document}